\documentclass[12pt]{article}

\font\grande=cmr9.5 scaled \magstep4
\font\medio=cmr9.5 scaled \magstep2
\outer\def\beginsection#1\par{\medbreak\bigskip
      \message{#1}\leftline{\bf#1}\nobreak\medskip
\vskip-\parskip
      \noindent}
\usepackage{graphicx} 
\begin{document}
\bibliographystyle {unsrt}

\titlepage

\begin{flushright}
CERN-PH-TH/2010-012
\end{flushright}

\vspace{15mm}
\begin{center}
{\grande Last scattering, relic gravitons}\\
\vspace{5mm}
{\grande and the circular polarization of the CMB}\\
\vspace{1.5cm}
 Massimo Giovannini 
 \footnote{Electronic address: massimo.giovannini@cern.ch} \\
\vspace{1cm}
{{\sl Department of Physics, 
Theory Division, CERN, 1211 Geneva 23, Switzerland }}\\
\vspace{0.5cm}
{{\sl INFN, Section of Milan-Bicocca, 20126 Milan, Italy}}
\vspace*{2cm}

\end{center}

\vskip 1cm
\centerline{\medio  Abstract}
The tensor contribution to the $V$-mode polarization induced by a magnetized plasma at last scattering vanishes exactly. Conversely a polarized background of relic gravitons cannot generate a $V$-mode polarization. The reported 
results suggest that, in the magnetized $\Lambda$CDM paradigm, the dominant 
source of circular dichroism stems from the large-scale fluctuations of the spatial curvature. 
\noindent

\vspace{5mm}

\vfill
\newpage
The linear polarization of the Cosmic Microwave Background radiation (CMB in what follows) is customarily 
understood in terms of a particular set of initial conditions of the radiative transfer equations stipulating that the fluctuations of the specific entropy are either absent or very small in comparison with the fluctuations 
of the spatial curvature over typical wavelengths larger than the Hubble radius at photon decoupling \cite{WMAP3}. 
In the current version  of the 
 $\Lambda$CDM paradigm\footnote{The acronym $\Lambda$CDM will be frequently used hereunder:
$\Lambda$ stands for the dark energy component and CDM stands for the cold dark matter component.}   the tensor modes of the geometry are absent. If included, they could also induce a purported tensor component of the B-mode power spectrum which has not yet been observed. Is it plausible that the CMB is also circularly polarized?  This occurrence cannot be a priori excluded. It would be highly desirable to have more direct upper limits on the circular polarizations  of the CMB with suitable low-frequency instruments \cite{tris1,tris2}  possibly operating in the GHz range or even below. When photons impinge electrons and ions in the presence of a magnetic field, the circular polarization arises naturally \cite{mg} leading to a non-vanishing $V$-mode whose magnitude depends upon the angular frequency of the experiment and upon the intensity of the radiation field. The $V$-mode polarization (as the intensity of the radiation field)  is invariant under two-dimensional rotations in the plane orthogonal to the direction of propagation of the 
photon \cite{chandra}. In \cite{mg}  only the  scalar mode contribution to the $V$-mode polarization has been considered and it 
is therefore natural to ask, in a complementary perspective,  if an additional source of circular polarization can arise from the tensor fluctuations of the geometry. In the tensor extension of the $\Lambda$CDM paradigm the long wavelength gravitons do contribute both to the linear polarization fluctuations as well as to the temperature fluctuations.  The magnetized electron-photon scattering induces a scalar $V$-mode polarization as well as a scalar intensity fluctuation \cite{mg};  the tensor modes are then expected to affect, in principle,  both the intensity
and the circular polarization. The aim of the present paper is an explicit calculation of the tensor contribution to the $V$-mode polarization in the framework of the magnetized $\Lambda$CDM paradigm. 
Consider the case of a conformally flat Friedmann-Robertson-Walker geometry, 
i.e. $\overline{g}_{\mu\nu}= a^2(\tau) \eta_{\mu\nu}$ where $\eta_{\mu\nu}$ is the Minkowski metric with signature $(+,\,-,\, -,\,-)$ and $a(\tau)$ is the scale factor.  The tensor fluctuations are defined as $\delta_{(\mathrm{t})}^{(1)} g_{i j} = -  a^2\,\, h_{ij}$
where $h_{i}^{i} = \partial_{i} h^{i}_{j} =0$ (Latin indices run over the three-dimensional spatial submanifold). The tensor fluctuation $h_{ij}(\vec{x},\tau)$ are 
 \begin{equation}
 h_{ij}(\vec{x},\tau) = \sum_{\lambda} h_{(\lambda)}(\vec{x},\tau) \epsilon_{ij}^{(\lambda)}(\hat{k}),\qquad \epsilon_{ij}^{(\lambda)} \epsilon_{ij}^{(\lambda')} = 2 \delta^{\lambda\lambda'},
 \label{SI4}
 \end{equation}
 where $\lambda = \oplus,\,\, \otimes$ denote the two polarizations; moreover 
$\epsilon_{ij}^{\oplus}(\hat{k}) = ( \hat{a}_{i} \hat{a}_{j} - \hat{b}_{i} \hat{b}_{j})$ 
and $\epsilon_{ij}^{\otimes}(\hat{k}) = (\hat{a}_{i} \hat{b}_{j} + \hat{a}_{j} \hat{b}_{i})$,
($\hat{a}$, $\hat{b}$ and $\hat{k} = \vec{k}/|\vec{k}|$ represent a triplet of mutually orthogonal unit vectors). The two circular polarizations of the graviton can also be defined as 
\begin{equation}
\epsilon_{ij}^{(\mathrm{L})}(\hat{k}) = \frac{1}{\sqrt{2}} [ \epsilon_{ij}^{\oplus}(\hat{k}) + i \epsilon_{ij}^{\otimes}(\hat{k})],\qquad \epsilon_{ij}^{(\mathrm{R})}(\hat{k})= \frac{1}{\sqrt{2}} [ \epsilon_{ij}^{\oplus}(\hat{k}) - i \epsilon_{ij}^{\otimes}(\hat{k})].
\label{GPOL}
\end{equation}
As the scalar modes of the geometry affect the evolution of the {\em scalar} contribution 
to the four Stokes parameters, the tensor modes of the geometry affect the evolution of the {\em tensor} component of the four Stokes parameters. The latter problem can be treated, in a 
magnetized environment, either through the Mueller or through the Jones calculus \cite{robson}. 
While the Mueller approach has been already employed in a related context \cite{mg}, 
the Jones method has the advantage of dealing directly with the components of the electric fields which are organized in a two-dimensional column vector. A hybrid approach will be employed hereunder: the Stokes parameters will appear as the components of a $2\times2$  polarization matrix, i.e. ${\mathcal P}$. The evolution 
of ${\mathcal P}$ can be formally written as 
\begin{equation}
 \frac{d {\mathcal P}}{d\tau} + \epsilon' {\mathcal P} = \frac{3 \epsilon'}{16 \pi} \int M(\Omega,\Omega')\, {\mathcal P}(\Omega,\Omega') \,M^{\dagger}(\Omega,\Omega')\, d\Omega',\qquad d\Omega' = d\cos{\vartheta'} \,d\varphi';
\label{N1}
\end{equation}
the dagger denotes the transposed and complex conjugate matrix;
$\epsilon' = x_{\mathrm{e}} \tilde{n}_{\mathrm{e}} \sigma_{\gamma\mathrm{e}} a(\tau)/a_{0}$ is the differential optical depth and $\sigma_{\gamma\mathrm{e}} = (8/3) \pi (e^2/m_{\mathrm{e}})^2$. The matrix ${\mathcal P}$ is
\begin{equation}
{\mathcal P}= \frac{1}{2} \left(\matrix{ I + Q
& U - i V &\cr
U + i V & I - Q &\cr}\right) = \frac{1}{2} \left( I\,{\bf 1} +U\, \sigma_{1} + V\, \sigma_2 +Q\, \sigma_3 \right),
\label{N2}
\end{equation}
where ${\bf 1}$ denotes the identity matrix while $\sigma_{1}$, $\sigma_{2}$ and $\sigma_{3}$ are the three 
Pauli matrices whose explicit expression allows for a swifter derivation of the collision term of Eq. (\ref{N1}). The primed angles denote conventionally the directions of the incident photons while the unprimed angles describe 
the scattered radiation field; with these specifications the matrix elements $M_{ij}$ (with 
$i,\, j = 1, 2$) is
\begin{eqnarray}
&& M_{11}\,= \zeta(\omega) \mu \mu' \Lambda_{1}(\omega) \cos{\Delta\varphi} 
- \sqrt{1 - \mu^2} \sqrt{1 - {\mu'}^2} \Lambda_{3}(\omega)
- i \Lambda_{2}(\omega) f_{\mathrm{e}}(\omega) \zeta(\omega) \mu\mu' \sin{\Delta\varphi}
\nonumber\\
&& M_{12}\, =  - \zeta(\omega) \mu \Lambda_{1}(\omega) \sin{\Delta\varphi} - i \Lambda_{2}(\omega) f_{\mathrm{e}}(\omega)
 \zeta(\omega) \mu \cos{\Delta\varphi},
\nonumber\\
&& M_{21}\, = \zeta(\omega) \mu' \Lambda_{1}(\omega)
 \sin{\Delta\varphi} + i f_{\mathrm{e}}(\omega) \Lambda_{2}(\omega) \zeta(\omega) \mu' \cos{\Delta\varphi},
\nonumber\\
&& M_{22}\, = \zeta(\omega) \Lambda_{1}(\omega) \cos{\Delta\varphi} 
- i f_{\mathrm{e}}(\omega) \Lambda_{2}(\omega) \zeta(\omega) \sin{\Delta\varphi}.
\label{mat}
\end{eqnarray}
where $\mu= \cos{\vartheta}$, $\mu' = \cos{\vartheta'}$ and $\Delta\varphi = (\varphi' - \varphi)$. The functions of the angular 
frequency of the photon $\omega = 2 \pi \nu$ appearing in Eq. (\ref{mat}) are: 
\begin{eqnarray}
&&\Lambda_{1}(\omega) = 1 + 
\biggl(\frac{\omega^2_{\mathrm{p\,\,i}}}{\omega^2_{\mathrm{p\,\,e}}}\biggr) \biggl( 
\frac{ \omega^2 - \omega^2_{\mathrm{B\,\,e}}}{\omega^2 - \omega^2_{\mathrm{B\,\,i}}}\biggr),\qquad \Lambda_{2}(\omega) = 1 - \biggl(\frac{\omega^2_{\mathrm{p\,\,i}}}{\omega^2_{\mathrm{p\,\,e}}}\biggr)
\biggl(\frac{\omega_{\mathrm{B\,\,i}}}{\omega_{\mathrm{B\,\,e}}}\biggr) \biggl( 
\frac{ \omega^2 - \omega_{\mathrm{B\,\,e}}^2}{\omega^2 - \omega^2_{\mathrm{B\,\,i}}}\biggr),
\label{LAM1}\\
&& \Lambda_{3}(\omega) = 1 + \biggl(\frac{\omega^2_{\mathrm{p\,\,i}}}{\omega^2_{\mathrm{p\,\,e}}}\biggr), \qquad 
f_{\mathrm{e}}(\omega) =  \biggl(\frac{\omega_{\mathrm{B\,\,e}}}{\omega}\biggr),
\qquad \zeta(\omega) = \frac{\omega^2}{\omega_{\mathrm{Be}}^2 - \omega^2}, 
\label{LAM3}
\end{eqnarray}
where $\omega_{\mathrm{B\,\,e,\,i}}$ and $\omega_{\mathrm{p\,\,e,\,i}}$ are the Larmor and plasma frequencies for electrons and ions depending, respectively, 
upon the magnetic field intensity and upon the relative charge concentrations. 
Equation (\ref{N1}) can be written, in components, as 
\begin{eqnarray}
&& \frac{d{\mathcal P}_{ij}}{d\tau} + \epsilon' {\mathcal P}_{ij} = \frac{3 \epsilon'}{16 \pi} \int  M_{i\,k}(\Omega,\Omega')\, 
 {\mathcal P}_{k\,m}(\Omega, \Omega')  \,M_{jk}^{*}(\Omega,\Omega')\, d\Omega',
\label{N2a}\\
&& \frac{d {\mathcal P}_{ij}}{d\tau}  =  \frac{\partial {\mathcal P}_{ij}}{\partial\tau} + n^{k} \, \frac{\partial {\mathcal P}_{ij}}{\partial x^{k}}  
- \frac{1}{2}  \frac{\partial h_{k\,m}}{\partial \tau} \,n^{k}\, n^{m} \frac{\partial {\mathcal P}_{ij}}{\partial \ln{q}},
\label{N2b}
\end{eqnarray}
where $q$ is the modulus of the (comoving) three-momentum.
The matrix elements ${\mathcal P}_{ij}$ can then be split as ${\mathcal P}_{ij} 
= f_{0}(q)[ \delta_{ij} + {\mathcal P}^{(1)}_{ij}]$ where $f_{0}(q)$ is the Bose-Einstein 
distribution as a function of the modulus of the comoving three-momentum $q$.
The matrix  ${\mathcal P}^{(1)}_{ij}$ contains the fluctuations of the brightness perturbations of the Stokes parameters whose explicit evolution is governed by the following set of equations:
\begin{eqnarray}
&& \frac{\partial \Delta^{(\mathrm{t})}_{\mathrm{I}}}{\partial \tau} + n^{k} \partial_{k} \Delta^{(\mathrm{t})}_{\mathrm{I}} +  \epsilon' \Delta^{(\mathrm{t})}_{\mathrm{I}} - \frac{\partial h_{k \, m}}{\partial \tau} n^{k} n^{m} =  \frac{3 \epsilon'}{16 \pi} \int_{-1}^{1} \,dx \int_{0}^{2\pi}\, dy\,\, {\mathcal C}_{\mathrm{I}}(\mu,x, \varphi,y),
\label{N4}\\
&&\frac{\partial \Delta^{(\mathrm{t})}_{\mathrm{Q}}}{\partial \tau} + n^{k} \partial_{k} \Delta^{(\mathrm{t})}_{\mathrm{Q}} + \epsilon' \Delta^{(\mathrm{t})}_{\mathrm{Q}} = \frac{3 \epsilon'}{16 \pi}
\int_{-1}^{1}\,dx \int_{0}^{2 \pi}\, dy\,\,{\mathcal C}_{\mathrm{Q}}(\mu, x,\varphi, y),
\label{N5}\\
&& \frac{\partial \Delta^{(\mathrm{t})}_{\mathrm{U}}}{\partial \tau} + n^{k} \partial_{k} \Delta^{(\mathrm{t})}_{\mathrm{U}} + \epsilon' \Delta^{(\mathrm{t})}_{\mathrm{U}} =  \frac{3 \epsilon'}{16 \pi}
\int_{-1}^{1} \,dx \int_{0}^{2\pi}\, dy\,\, {\mathcal C}_{\mathrm{U}}(\mu,x,\varphi,y),
\label{N6}\\
&& \frac{\partial \Delta^{(\mathrm{t})}_{\mathrm{V}}}{\partial \tau} + n^{k} \partial_{k} \Delta^{(\mathrm{t})}_{\mathrm{V}}+ \epsilon' \Delta^{(\mathrm{t})}_{\mathrm{V}}  =  \frac{3 \epsilon'}{16 \pi}
\int_{-1}^{1} \,dx \int_{0}^{2\pi}\, dy\,\,{\mathcal C}_{\mathrm{V}}(\mu,x,\varphi,y).
\label{N7}
\end{eqnarray}
For sake of conciseness\footnote{It is understood that the 
$\Delta^{(\mathrm{t})}_{\mathrm{I}}$, $\Delta^{(\mathrm{t})}_{\mathrm{Q}}$, $\Delta^{(\mathrm{t})}_{\mathrm{U}}$ and 
$\Delta^{(\mathrm{t})}_{\mathrm{V}}$ appearing in the source functions at the right hand side 
of Eqs. (\ref{N4})--(\ref{N7}) do all depend upon $(x,y)$; it is also understood 
that all the coefficients $C_{i}^{j}$ of Eqs. (\ref{N8})--(\ref{N11}) do depend upon 
$(\mu,x, \varphi,y)$.} the functional dependence can be dropped
\begin{eqnarray}
 {\mathcal C}_{\mathrm{I}}(\mu,x, \varphi,y) &=& C_{\mathrm{I}}^{\mathrm{I}} \Delta^{(\mathrm{t})}_{\mathrm{I}} + C_{\mathrm{I}}^{\mathrm{Q}} \Delta^{(\mathrm{t})}_{\mathrm{Q}}
+ C_{\mathrm{I}}^{\mathrm{U}}\Delta^{(\mathrm{t})}_{\mathrm{U}} + C_{\mathrm{I}}^{\mathrm{V}} \Delta^{(\mathrm{t})}_{\mathrm{V}},
\label{N8}\\
 {\mathcal C}_{\mathrm{Q}}(\mu,x, \varphi,y) &=& C_{\mathrm{Q}}^{\mathrm{I}} \Delta^{(\mathrm{t})}_{\mathrm{I}}
+ C_{\mathrm{Q}}^{\mathrm{Q}} \Delta^{(\mathrm{t})}_{\mathrm{Q}} 
+ C_{\mathrm{Q}}^{\mathrm{U}} \Delta^{(\mathrm{t})}_{\mathrm{U}}+ C_{\mathrm{Q}}^{\mathrm{V}}\Delta^{(\mathrm{t})}_{\mathrm{V}},
\label{N9}\\
 {\mathcal C}_{\mathrm{U}}(\mu,x, \varphi,y) &=& C_{\mathrm{U}}^{\mathrm{I}}( \Delta^{(\mathrm{t})}_{\mathrm{I}} + C_{\mathrm{U}}^{\mathrm{Q}} \Delta^{(\mathrm{t})}_{\mathrm{Q}}
+ C_{\mathrm{U}}^{\mathrm{U}} \Delta^{(\mathrm{t})}_{\mathrm{U}} + C_{\mathrm{U}}^{\mathrm{V}}\Delta^{(\mathrm{t})}_{\mathrm{V}},
\label{N10}\\
{\mathcal C}_{\mathrm{V}}(\mu,x, \varphi,y) &=& C_{\mathrm{V}}^{\mathrm{I}} \Delta^{(\mathrm{t})}_{\mathrm{I}}+ C_{\mathrm{V}}^{\mathrm{Q}} \Delta^{(\mathrm{t})}_{\mathrm{Q}} 
+ C_{\mathrm{V}}^{\mathrm{U}} \Delta^{(\mathrm{t})}_{\mathrm{U}} + C_{\mathrm{V}}^{\mathrm{V}} \Delta^{(\mathrm{t})}_{\mathrm{V}}.
\label{N11}
\end{eqnarray}
The superscript $(\mathrm{t})$ on the brightness perturbations 
signifies that we are dealing with the tensor modes: similar equations can be 
written in the case of the scalar modes. The coefficients appearing in Eqs. (\ref{N8}) 
and (\ref{N11}) are given by:
\begin{eqnarray}
C_{\mathrm{I}}^{\mathrm{I}} &=&  \Lambda_{3}^{2}{\mathcal A}^2(\mu,x) - 2 \zeta\, \Lambda_{1}\, \Lambda_{3}\, \mu \,x \,{\mathcal A}(\mu,x)\,c(y,\varphi)  
\nonumber\\
&+& \zeta^2 [ f_{\mathrm{e}}^2 \Lambda_{2}^2 ( \mu^2 + x^2) + \Lambda_{1}^2 ( 1+ x^2 \mu^2)] c^2(y,\varphi)
\nonumber\\
&+&  
\zeta^2 [  \Lambda_{1}^2 ( \mu^2 + x^2) + f_{\mathrm{e}}^2\Lambda_{2}^2 ( 1+ x^2 \mu^2)] s^2(y,\varphi), 
\nonumber\\
C_{\mathrm{I}}^{\mathrm{Q}} &=& \Lambda_{3}^{2}{\mathcal A}^2(\mu,x) - 2 \zeta \Lambda_{1}\Lambda_{3}\,\mu\,x\, {\mathcal A}(\mu,x)\,c(y,\varphi)  
\nonumber\\
&+& \zeta^2 [ f_{\mathrm{e}}^2 \Lambda_{2}^2 (- \mu^2 + x^2) + \Lambda_{1}^2 ( - 1+ x^2 \mu^2)] c^2(y,\varphi) 
\nonumber\\
&+&  \zeta^2 [  \Lambda_{1}^2 ( -\mu^2 + x^2) + f_{\mathrm{e}}^2\Lambda_{2}^2 (- 1+ x^2 \mu^2)] s^2(y,\varphi), 
\nonumber\\
C_{\mathrm{I}}^{\mathrm{U}} &=& 2 \zeta [ \Lambda_{1}\, \Lambda_{3}\, \mu \,{\mathcal A}(\mu,x) - \zeta ( \Lambda_{1}^2 - f_{\mathrm{e}}^2 \Lambda_{2}^2)( \mu^2 -1) \,x\, 
c(y,\varphi)] s(y,\varphi),
\nonumber\\
C_{\mathrm{I}}^{\mathrm{V}} &=& 2 f_{\mathrm{e}} \zeta \Lambda_{2} 
[ \zeta \Lambda_{1} ( 1 + \mu^2) x - \Lambda_{3} \mu {\mathcal A}(\mu,x)\, c(y,\varphi)]
\nonumber\\
C_{\mathrm{Q}}^{\mathrm{I}} &=&  \Lambda_{3}^{2}{\mathcal A}^2(\mu,x) - 2 \zeta \Lambda_{1}\Lambda_{3} \mu\,
x\,{\mathcal A}(\mu,x)\,c(y,\varphi) 
\nonumber\\
&+& \zeta^2 [ f_{\mathrm{e}}^2 \Lambda_{2}^2 ( \mu^2 - x^2) + \Lambda_{1}^2 (- 1+ x^2 \mu^2)]c^2(y,\varphi)
\nonumber\\
 &+&  
\zeta^2 [  \Lambda_{1}^2 ( \mu^2 - x^2) + f_{\mathrm{e}}^2\Lambda_{2}^2 ( -1+ x^2 \mu^2)]s^2(y,\varphi),
\nonumber\\
C_{\mathrm{Q}}^{\mathrm{Q}} &=&  \Lambda_{3}^{2}{\mathcal A}^2(\mu,x) - 2 \zeta \Lambda_{1}\Lambda_{3}\, \mu\,
x \,{\mathcal A}(\mu,x) \,c(y,\varphi) 
\nonumber\\
&+& \zeta^2 [ -f_{\mathrm{e}}^2 \Lambda_{2}^2 ( \mu^2 + x^2) + \Lambda_{1}^2 (1+ x^2 \mu^2)]\,c^2(y,\varphi) 
\nonumber\\
&+&  
\zeta^2 [  -\Lambda_{1}^2 ( \mu^2 + x^2) + f_{\mathrm{e}}^2\Lambda_{2}^2 ( 1+ x^2 \mu^2)]s^2(y,\varphi) ,
\nonumber\\
C_{\mathrm{Q}}^{\mathrm{U}} &=& 2 \zeta [ \Lambda_{1} \Lambda_{3} \mu {\mathcal A}(\mu,x) - \zeta ( \Lambda_{1}^2 - f_{\mathrm{e}}^2 \Lambda_{2}^2)( \mu^2 +1) x c(y,\varphi)]s(y, \varphi),
\nonumber\\
C_{\mathrm{Q}}^{\mathrm{V}} &=& 2 f_{\mathrm{e}} \zeta \Lambda_{2} [ \zeta \Lambda_{1} ( -1 + \mu^2) x - \Lambda_{3} \mu 
{\mathcal A}(\mu,x)c(y,\varphi),
\nonumber\\
C_{\mathrm{U}}^{\mathrm{I}} &=& 2 \zeta [ - \Lambda_{1}  \Lambda_{3} x {\mathcal A}(\mu,x) + \zeta ( \Lambda_{1}^2 - f_{\mathrm{e}}^2 \Lambda_{2}^2) \mu ( x^2 -1) 
c(y,\varphi)] s(y,\varphi),
\nonumber\\
C_{\mathrm{U}}^{\mathrm{Q}} &=& 2 \zeta [ - \Lambda_{1}  \Lambda_{3} x {\mathcal A}(\mu,x)  + \zeta ( \Lambda_{1}^2 - f_{\mathrm{e}}^2 \Lambda_{2}^2) \mu ( x^2 +1)c(y, \varphi)] s(y,\varphi),
\nonumber\\
C_{\mathrm{U}}^{\mathrm{U}} &=& - 2 \zeta [ \Lambda_{1} \Lambda_{3} {\mathcal A}(\mu,x)c(y,\varphi) - 
\zeta (\Lambda_{1}^2 - f_{\mathrm{e}}^2 \Lambda_{2}^2 )\,\mu\, x\, (c^2(y,\varphi) - s^2(y,\varphi))],
\nonumber\\
C_{\mathrm{U}}^{\mathrm{V}} &=& - 2 f_{\mathrm{e}} \zeta \Lambda_{2} \Lambda_{3} {\mathcal A}(\mu,x),
\nonumber\\
C_{\mathrm{V}}^{\mathrm{I}} &=& 2 f_{\mathrm{e}} \zeta \Lambda_{2} [ \zeta \Lambda_{1} \mu ( 1 + x^2 ) - \Lambda_{3}\,{\mathcal A}^2(\mu,x)\, x \,c(y,\varphi)],
\nonumber\\
C_{\mathrm{V}}^{\mathrm{Q}} &=& 2 f_{\mathrm{e}} \zeta \Lambda_{2} [ \zeta \Lambda_{1} \mu ( -1 + x^2 ) - \Lambda_{3} 
{\mathcal A}(\mu,x) \,x\, c(y,\varphi)],
\nonumber\\
C_{\mathrm{V}}^{\mathrm{U}} &=& 2 f_{\mathrm{e}} \zeta \Lambda_{2} \Lambda_{3} {\mathcal A}(\mu,x) s(y,\varphi)
\nonumber\\
C_{\mathrm{V}}^{\mathrm{V}} &=& 2 i \zeta [ \zeta ( \Lambda_{1}^2 + f_{\mathrm{e}}^2 \Lambda_{2}^2) \mu x - \Lambda_{1} \Lambda_{3} {\mathcal A}(\mu,x) c(y, \varphi)];
\label{N12}
\end{eqnarray}
in Eq. (\ref{N12}) the shorthand notations ${\mathcal A}(\mu,x) = \sqrt{1 - \mu^2}\sqrt{1 - x^2}$, $c(y,\varphi) = \cos{(y - \varphi)}$ and $s(y,\varphi) = \sin{(y - \varphi)}$
have been introduced. 
We shall now prove that, in spite of the polarization of the graviton,
the circular polarization will not be induced at last scattering even in the presence of 
a very strong magnetic field. The obtained equations generalize 
the standard system obtained (in the absence of magnetic fields) for the 
tensor components of the Stokes parameters (see, for instance, \cite{gweq}).
Consider, for sake of concreteness, a linearly polarized graviton 
(for instance $\oplus$). Assuming, without loss of generality, that the graviton travels along the $\hat{z}$ axis, the combination $h_{k\,m} n^{k} n^{m}$ equals $h_{\oplus} ( 1 - \mu^2) \cos{(2 \varphi)}$. 
The dependence of the brightness perturbations upon $\varphi$ can then be deduced 
from the whole symmetry of the system and it is\footnote{We shall not be directly interested here in the space-time dynamics of the relic gravitons but rather on their polarization properties; different background models lead to different frequency dependence of the relic graviton backgrounds (see, for instance, \cite{mg2}); these 
aspects will not be directly relevant for the present considerations.} 
\begin{eqnarray}
&&\Delta^{(\mathrm{t})}_{\mathrm{I}}(x,y) = {\mathcal Z}( 1 - x^2) \cos{2 y}, 
\qquad \Delta^{(\mathrm{t})}_{\mathrm{Q}}(x,y) = {\mathcal T} \, ( 1 + x^2) \,\cos{2y}, 
\nonumber\\
&&\Delta^{(\mathrm{t})}_{\mathrm{U}}(x,y) = - 2 x\, {\mathcal T}\,\sin{2 y}, \qquad 
\Delta^{(\mathrm{t})}_{\mathrm{V}}(x,y) = 2\, x\, {\mathcal S}\, \cos{2 y}.
\label{ans1}
\end{eqnarray}
Using Eq. (\ref{ans1}) into Eqs. (\ref{N7})--(\ref{N11}) and taking into account 
Eq. (\ref{N12}) the evolution of  ${\mathcal Z}$, ${\mathcal T}$ and ${\mathcal S}$ become, after lengthy algebra
\begin{eqnarray}
&& \frac{\partial {\mathcal Z}}{\partial\tau} + n^{k} \partial_{k} {\mathcal Z} + \epsilon' {\mathcal Z} - 
\frac{\partial h_{\oplus}}{\partial \tau}   =  \epsilon' \zeta^2(\omega) [\Lambda_{1}^2(\omega) - f_{\mathrm{e}}^2(\omega) \Lambda_{2}^2(\omega)] \Sigma^{(\mathrm{t})},
\label{N30}\\
&& \frac{\partial {\mathcal T}}{\partial\tau} +  n^{k} \partial_{k} {\mathcal T} + \epsilon' {\mathcal T} = -  \epsilon' \zeta^2(\omega) [\Lambda_{1}^2(\omega) - f_{\mathrm{e}}^2(\omega) \Lambda_{2}^2(\omega)] \Sigma^{(\mathrm{t})},
\label{N31}\\
&& \frac{\partial {\mathcal S}}{\partial\tau} +  n^{k} \partial_{k} {\mathcal S} + \epsilon' {\mathcal S}=0,
\label{N32}
\end{eqnarray}
where, defining $\int_{-1}^{1} {\mathcal Z} P_{\ell} dx = 2 (-i)^{\ell} {\mathcal Z}_{\ell}$ and 
$\int_{-1}^{1} {\mathcal T} P_{\ell} dx = 2 (-i)^{\ell} {\mathcal T}_{\ell}$, the source term $\Sigma^{(\mathrm{t})}$ can also be expressed as
\begin{eqnarray}
\Sigma^{(\mathrm{t})} &=& \frac{3}{32} \int_{-1}^{1} d x 
[  (1 - x^2)^2 {\mathcal Z}(x)
- ( 1 + x^2)^2 {\mathcal T}(x) - 4 x^2 {\mathcal T}(x)] 
\nonumber\\
&=& \frac{3}{70}{\mathcal  Z}_{4} + \frac{{\mathcal Z}_{2}}{7} - \frac{{\mathcal Z}_{0}}{10}- \frac{3}{70} {\mathcal T}_{4} + \frac{6}{7} {\mathcal T}_{2} - \frac{3}{5} {\mathcal T}_{0}.
\label{MM}
\end{eqnarray}
The integrations over $y$ involves simple trigonometric identities:
\begin{eqnarray} 
&& \int_{0}^{2\pi} \cos{2(\varphi' - \varphi)} \, \cos{2\varphi'}\, d\varphi' = 
\int_{0}^{2\pi} \sin{2(\varphi' - \varphi)} \, \sin{2\varphi'}\, d\varphi' = \pi \cos{2\varphi},
\label{int1}\\
&& \int_{0}^{2\pi} \sin{2(\varphi' - \varphi)} \, \cos{2\varphi'}\, d\varphi' = -\pi \sin{2 \varphi} = -  \int_{0}^{2\pi} \cos{2(\varphi' - \varphi)} \, \sin{2\varphi'}\, d\varphi'.
\label{int2}
\end{eqnarray}
In the case of the orthogonal polarization of the graviton (i.e. $\otimes$) the combination 
$h_{k\,m} n^{k} n^{m}$ equals $h_{\otimes} ( 1 - \mu^2) \sin{2 \varphi}$ and symmetry considerations imply that 
\begin{eqnarray}
&&\Delta^{(\mathrm{t})}_{\mathrm{I}}(x,y) = {\mathcal Z}\,( 1 - x^2)\,\sin{2 y}, 
\qquad \Delta^{(\mathrm{t})}_{\mathrm{Q}}(x,y) = {\mathcal T}\,( 1 + x^2)\,\sin{2y}, 
\nonumber\\
&&\Delta^{(\mathrm{t})}_{\mathrm{U}}(x,y) =  2\,x\,{\mathcal T}\,\cos{2 y}, \qquad 
\Delta^{(\mathrm{t})}_{\mathrm{V}}(x,y) = 2\,x\,{\mathcal S}\,\sin{2 y}.
\label{N33} 
\end{eqnarray}
Inserting Eq. (\ref{N33}) inside Eqs. (\ref{N7})--(\ref{N11}) and performing the 
integration over $y$ the same results of Eqs. (\ref{N31})--(\ref{N33}) can be obtained.
The same discussion can be carried on when the gravitons are circularly polarized since the left and right polarizations can be expressed as  linear 
combinations of $\oplus$ and $\otimes$ (see Eq. (\ref{GPOL})).  
It should be remarked that  
magnetic fields can also induce relic gravitons on their own (see, e.g. \cite{gwm1})
but the primary goal of this paper is to discuss the interplay of circular dichroism 
of the CMB and the tensor modes of the geometry. It could be swiftly argued, a posteriori, that the rationale for the obtained result is just a consequence 
of the transformation properties of $\Delta_{\mathrm{V}}(\hat{n}) $ which is a scalar 
under two dimensional rotations in the plane orthogonal to $\hat{n}$. The latter 
argument can be immediately refuted since also $\Delta_{\mathrm{I}}(\hat{n})$ transforms like a scalar for rotations in the plane orthogonal to $\hat{n}$ but, nonetheless,  it is affected by the tensor modes of the geometry even in the absence 
of magnetic fields as it is clear from Eq. (\ref{N30}) (see also \cite{gweq}). 
Equations (\ref{N31})--(\ref{N33}) can be usefully contrasted with their scalar 
counterpart in the limit of $f_{\mathrm{e}}(\omega) < 1$ \cite{mg} 
\begin{eqnarray} 
&& \frac{\partial \Delta^{(\mathrm{s})}_{\mathrm{I}}}{\partial \tau} + n^{k} \partial_{k}(\Delta^{(\mathrm{s})}_{\mathrm{I}} + \phi) + \epsilon' \Delta^{(\mathrm{s})}_{\mathrm{I}} = \psi' + 
\epsilon'\biggl[ \mu v_{\mathrm{b}} + \Delta^{(\mathrm{s})}_{\mathrm{I}0} - \frac{P_{2}(\mu) }{2} S_{\mathrm{Q}}\biggr],
\label{N40a}\\
&&  \frac{\partial \Delta^{(\mathrm{s})}_{\mathrm{Q}}}{\partial\tau} + n^{k} \partial_{k}\Delta^{(\mathrm{s})}_{\mathrm{Q}}  + \epsilon' \Delta^{(\mathrm{s})}_{\mathrm{Q}} = 
\frac{3(1- \mu^2)\epsilon'}{4} S_{\mathrm{Q}},
\label{N40b}\\
&& \frac{\partial \Delta^{(\mathrm{s})}_{\mathrm{V}}}{\partial \tau} + n^{k} \partial_{k} 
\Delta^{(\mathrm{s})}_{\mathrm{V}} 
= \epsilon' \biggl\{ f_{\mathrm{e}}(\omega) [ 2 \Delta^{(\mathrm{s})}_{\mathrm{I}0} - S_{\mathrm{Q}}\biggr] - \frac{3}{4} i \Delta^{(\mathrm{s})}_{\mathrm{V}1}\biggr\},
\label{N40}
\end{eqnarray}
where the subscript $(\mathrm{s})$ reminds that we are dealing with the scalar 
modes; as usual $S_{\mathrm{Q}} =( \Delta^{(\mathrm{s})}_{\mathrm{I}2} + \Delta^{(\mathrm{s})}_{\mathrm{Q}0} 
+ \Delta^{(\mathrm{s})}_{\mathrm{Q}2})$ and $\Delta^{(\mathrm{s})}_{\mathrm{U}} =0$.
Equation (\ref{N40}) possesses a source term inducing a computable amount of circular polarization at last scattering. 
Conversely Eq. (\ref{N32}) does not have a source term. This shows 
that, in spite of the polarization of the graviton, the presence 
of a magnetic field at last scattering does not induce a tensor 
contribution to the $V$-mode polarization. The latter result does not 
forbid the presence of a tensor contribution to the $V$-mode polarization 
stemming from the initial conditions (i.e. if circular polarization was already 
present prior to last scattering). However, if the CMB is assumed 
to be unpolarized before last scattering we can conclude that a computable 
amount of circular polarization can be induced in the presence of a magnetic field 
only from the scalar modes of the geometry and not from the tensor modes. 
The author is indebted to Giorgio Sironi for stimulating conversations.

\end{document}